# Ultrafast reprogrammable multifunctional vanadium-dioxide-assisted metasurface for dynamic THz wavefront engineering

Javad Shabanpour *, Sina Beyraghi, Ahmad Cheldavi

Department of Electrical Engineering, Iran University of Science and Technology, Narmak, Tehran 16486-13114, Iran, m_shabanpour@elec.iust.ac.ir, m_beyraghi @elec.iust.ac.ir, cheldavi@iust.ac.ir

Corresponding Author: m_shabanpour@elec.iust.ac.ir

*Abstract*

In this paper, for the first time, a new generation of ultrafast reprogrammable multi-mission bias encoded metasurface is proposed for dynamic THz wavefront engineering by employing VO2 reversible and fast monoclinic to tetragonal phase transition. The multi-functionality of our designed VO2 based coding metasurface (VBCM) was guaranteed by elaborately designed meta-atom comprising three-patterned VO2 thin films whose operational statuses can be dynamically tuned among four states of "00"- "11" by merely changing the biasing voltage controlled by an external FPGA platform. Capitalizing on such meta-atom design and by driving VBCM with different spiral-like and spiral-parabola-like coding sequences, single vortex beam and focused vortex beam with interchangeable OAM modes were satisfactorily generated respectively. Additionally, by adopting superposition theorem and convolution operation, symmetric/asymmetric multiple beams and arbitrarily-oriented multiple vortex beams in pre-demined directions with different topological charges are realized. The versatility of our designed VBCM also has equipped a platform to focus the incident THz wavefront into a pre-determined point which can be dynamically altered. Several illustrative examples successfully have clarified that proposed VBCM is a promising candidate for solving crucial THz challenges such as high data rate wireless communication where ultrafast switching between several missions is required.

*1.Introduction*

In recent years, the scope of THz science and technologies has reached a maturity and attracted massive attention due to their potential applications like biomedicine [1], security checking and high data rate transmit through wireless communication [2-4]. However, as the key technique, manipulating EM waves reveals the necessity of employing metasurfaces as the two-dimensional analogous of more general volumetric metamaterials to furnish an inspiring groundwork for realizing some rich diverse applications such as, but not limited to, invisibility cloaks [5,6], negative refraction [7], optical illusion [8] or epsilon near zero behaviors [9,10].

Recently Giovampaola and Engheta have brought up the subject of "Digital Metamaterials" in 2014 [11]. In an effort to manipulate the EM waves with more degrees of freedom, this simple, yet powerful concept led to formation the idea of "Coding Metasurface" in the same year [12]. Due to the lower weight and being easier to design and fabricate, digital metasurfaces have experienced rapid development compared to traditional wave manipulation [13]. By purposefully distributing coding particles over a 2D plane in a periodic or aperiodic manner, a variety of exquisite physics phenomena and innovative EM devices have been created [14-18].

However, in most of these strategies, the metasurfaces are designed for a specific application and their functionalities remain fixed once they are constructed. For example, Shao *et al*. proposed a dielectric 2-bit coding metasurface with distinct functionalities from anomalouse reflection to vortex beam generation [19]. Dual-band 2-bit coding metasurface to fully control vortex beams carrying orbital angular momentum with different topological charges (*L*) was discussed in [20], But the lack of adjustability significantly hinders their practical applications.

Owing to the increasing need for system integration, a single metasurface that possesses multiple diversified functionalities in a real time manner with tunable meta-atoms is highly desired notably at THz frequencies. [21-23]. Until the present, some papers have brought forward electronically tunable coding metasurface in microwave frequency by using pin diodes in each coding elements [24-26]. For example, Huang *et al.* presented a method to design 2-bit digitally-controlled coding metasurface in order to realize different far-field patterns. By integrating two pin diodes in each coding particles and switching their operating states, producing four phase responses was realized [27]. In the light of the complexity and lack of commercially access of active elements (pin or varactor diodes) at high frequencies, scaling these devices to THz regime is very difficult if not impossible [28].

For real-time manipulating THz waves, an efficient technique is to integrate standard metasurface with phase-change materials (PCM), for instance, graphene [29-31], Liquid crystal [32-35] and vanadium dioxide (VO2) [36-38]. To implement such a platform, we have benefited from VO2 exotic properties. VO2 is a smart material, that undergoes an ultrafast and brutal reversible first-order phase transition from insulating monoclinic (P2$_1$/c) to the metallic tetragonal (P4$_2$/mnm) phase, above critical temperature $T_C = 68^0 C$ [38]. Since the critical temperature is a function of V-V distance in crystal structure, benefiting from high pressure and doping technique, decreasing in critical temperature would be feasible [92,93]. This metal-insulator transition (MIT) can be provoked by thermal [39,40], optical [41-43] or electrical (charge injection or Joule heating) [44-50] stimuli. Such a transition is due to the physical structure is still under debate [51-53]. This transition in VO2 can occur within an order of several nanoseconds or even in picoseconds range for optical activation [54]. The VO2 phase alternation that was studied in [55] and [56] through time-resolved x-ray diffraction and time-resolved optical transmission respectively, revealed that only short time (<500 fs) was required for taking place the MIT. Moreover, Cavalleri's study indicated that the photo-induced transition time for VO2 thin film (50nm) could reach 80 fs [57].

Electrical and optical property of VO2 has dramatic change (4 to 5 order of magnitude change of the electrical conductivity) across the two phases [36,]. Owing to the ultrafast transition duration, almost near room critical temperature and fruitful structural transition behaviors, VO2 becomes a striking material in tunable metamaterial devices at GHz [58,59], optics [60-62] and has numerous fantastic applications in THz frequencies such as reconfigurable THz filters [63,64], polarization converter [65-69], reflection/transmission THz waves modulator [70-72], tunable THz absorbers [73,74] and reconfigurable antennas [75,76]. In 2016, a simple VO2-assisted digital metasurface was proposed to dynamically control the near infrared light [77]. By allowing voltage to be locally applied to the VO2, and distributing unit cells in one direction, switchable beam splitters with only limited splitting angles have been realized. Regarding that the reflection phase of the proposed structure is either 0° or 180° relying on the state of VO2, several basic functionalities and fundamental THz challenges such as anomalous reflection, asymmetric multi-beams and vortex beam generation with adjustable properties are not accessible with this structure.

In departure from the abovementioned work, wherein the mission of utilizing VO2 in the structure is limited to only a single tunable function, we present, to the best of author's knowledge, the first VO2-based coding metasurface (VBCM) that can be reprogrammed for realizing multi-type functionalities from vortex beams with different topological charges toward emitting multiple arbitrarily-oriented pencil beams. Although, some articles [78,79] have reported graphene-based coding metasurface for real-time manipulation of THz waves recently, but the graphene is of great loss compared to VO2 [80] such as difficult fabrication process as respects that the graphene consists only of a monolayer of carbon atoms compared to volumetric VO2 [77] and very low switching time (thousands times slower than VO2). Although today's applications in THz regime as high data rate wireless communication and ultra-massive MIMO communication are in dire need of ultrafast THz wavefront manipulation, we are not aware of any reports of ultrafast real-time THz wavefront engineering and this field is still largely unexplored and we believe that our proposed ultrafast VBCM structure has the great potentials to fill this gap.

In this paper, an ultrafast versatile 2-bit VO2-based coding metasurface with programmable meta-atoms is designed whose operational statuses can be dynamically switched between four states of "00", "01", "10", and "11". By applying external bias voltage to the elaborately designed meta-atom controlled by an FPGA platform, the capability of manipulating THz wavefront in a real-time manner has been achieved. Each constitutive unitcell of the VBCM integrates three patterned VO2 layers fed by two biasing voltages (ON/OFF), therefore it can alterably possess four reflection phase responses of $0$, $\pi/2$, $\pi$, and $3\pi/2$ without changing the geometrical parameters. In order to expose the capacities of our structure in multifunctional wavefront engineering, we have arranged various coding samples to accomplish different interchangeable functions from vortex beams generation with different OAM modes toward emitting multiple arbitrarily-oriented pencil beams without re-optimizing or re-fabricating structure. Benefited from convolutional and superposition theorems on the far-field pattern, several THz fundamental challenges such as symmetric/asymmetric multiple pencil beams and arbitrarily-oriented multiple vortex beams with different OAM modes, with ultrafast switching time are solved by our proposed VBCM structure. As proof of concept, number of illustrative examples validated through numerical simulations and theoretical predictions. To the best of our insight, this is the first ultrafast re-programmable coding metasurface based on VO2 phase transition that armed a platform for realizing some rich interchangeable missions at THz frequencies. The authors believe that the proposed VBCM paves the way for ultrafast multifunctional THz wavefront engineering and future practical applications.

*2.Design of the switchable multifunctional VBCM*

Fig 1. represents the basic meta-atom of the VO2 integrated metasurface composing of three layers, which, from top to bottom, are the VO2 layers, dielectric substrate, and a gold plane to impede the transmission energy into the back of the VBCM. At the top layer, three patterned VO2 layers deposited on the sapphire as the suitable substrate ($\varepsilon_r = 9 \cdot 4. tan\delta = 0 \cdot 0001$) with the thickness of h = 38μm. Up to now, VO2 thin films have been prepared on different substrates as silica glass [81], polyimide [80,66], silicon [82-84] and sapphire [85,86,63] by several techniques from chemical vapor deposition [87] to reactive electron-beam evaporation [88] but it is preferable to use sapphire substrate to reach high quality VO2 thin films due to the beneficial lattice matching effect [89].

The complex dielectric properties of the VO2 can be characterized by the Bruggeman effective-medium theory in the THz range, wherein, $\varepsilon_d$ and $\varepsilon_m$ denote the dielectric constant of the semiconductor and metallic regions respectively and V represents the volume fraction of metallic regions [65].

$$\varepsilon_{VO2} = \frac{1}{4}\{\varepsilon_d(2-3V) + \varepsilon_m(3V-1) + \sqrt{[\varepsilon_d(2-3V) + \varepsilon_m(3V-1)]^2 + 8\varepsilon_d\varepsilon_m}\} \quad (1)$$

At room temperature the dielectric constant of VO2 is about 9 in the insulting state [66,69,90,91] and by applying external bias voltage directly to VO2 thin film, the structural transformation occurs and VO2 turns into the rutile phase. Typical VO2 films that are grown on c-type or r-type sapphire substrate, display electrical conductivity in the range of $10 \sim 100 \; S/m$ in insulating state and as high as an order of $10^5 \; S/m$ in the metallic state [63]. Frequency independent conductivity of VO2 is set to be $\sigma = 10 \; S/m$ (OFF state) [81,91] and $\sigma = 5 \times 10^5 \; S/m$ (ON state) [45,67,94] in the insulating and metallic phase corresponding to $T_C = 300K$ and $T_H = 400K$ respectively. The periodicity of our subwavelength meta-atoms is $P = 100$μm. The other geometrical parameters are $w_1 = 90$μm. $w_2 = 62$μm. $w_3 = 14$μm. $a_1 = 39$μm. $a_2 = 26$μm. $b_1 = 90$μm. $b_2 = 52$μm. $b_3 = 19$μm. $c_1 = 14$μm. $c_2 = 23$μm, respectively. The thickness of VO2 film is assumed to be $t = 1$μm. All of these geometrical dimensions are extracted from an extensive simulation and optimization process to obtain four reflection phases of 0, $\pi/2$, $\pi$, and $3\pi/2$ (phase step of 90°) to mimic four digital states of "00", "01", "10", "11" and any change in the dimensions lead to deteriorate this phase differences. Contrary to geometrically encoded metasurface, the employed meta-atoms of the VBCM structure in this study have the same dimensions but with different properly bias voltages leading to call this structure as bias-encoded metasurface. All the numerical simulations are carried out by means of Commercial software CST Microwave Studio. For evaluating the reflection characteristics for the infinite array of digital elements, open (add space) boundary condition is applied along z-axis whilst periodic boundary conditions are utilized along the x and y directions to incorporate the mutual coupling effect between neighboring elements. The simulated reflection spectra are depicted in Fig. 2 for different sets of external bias voltage. It is clearly observed from Fig. 2 that with appropriately biasing three patterned VO2 layers from top to bottom as follows: ("OFF/OFF/OFF", "ON/ON/ON", "ON/OFF/ON" and "OFF/ON/OFF") corresponding to four digital states of "00", "01", "10", "11" respectively, our elaborately designed meta-particle has been successful in providing a phase step of 90° with reflection amplitudes above 0.82 at 0.44 THz which means that the meta-atom non-absorptive behavior has been achieved. We have deliberately selected these two temperature points since VO2 at these two points is in the dielectric or metallic steady state phases [77]. Furthermore, the Ohmic loss of the meta-atom is maximum at intermediate temperatures which lead to a sharp drop in the reflection amplitude which makes device efficiency to be minimized. It is worth mentioning that to maintain the symmetry of the simulation results, both top and bottom VO2 layers (red colors in Fig. 1) must be geometrically identical and switched "ON" and "OFF" simultaneously. Finally considering the abovementioned constraints and avoiding VO2 intermediate phases, employing three patterned VO2 layers in the meta-particles of the 2-bit VBCM structure that leads to the complexity of the biasing system is forced to users. It can be found from Fig. 2 that the perfect 2-bit characteristic occurs at 0.44 THz which means that the approximate size of our subwavelength meta-atoms equals to $\lambda/7$. Meanwhile, a maximum phase range of 260° is attained here that is wide enough for acceptable functioning of our ultrafast multifunctional VBCM for real-time THz wavefront engineering.

Before beginning next section, it is worthwhile mentioning that the multi-functionality of VBCM structure can be realized when external dc bias voltage/current is applied to each individual unit-cell independently through the Au electrodes deposited on the top of the patterned VO2 layers to form a good Ohmic contact. The main mechanism for this voltage/current driven IMT process remains controversial as it may originate from Joule heating [44,96] or electric field effects

[95,97]. Recently the viewpoint of zimmers *et al* [44], was that the local Joule heating plays a predominant role in the dc voltage (or dc current) induced IMT. Wu et al [95] claim that Joule heating effect was negligible and electric field alone is sufficient to induce the IMT. Beyond these two competing claims, some papers were commented that these two main mechanism are considered to be mixed together and were difficult to disentangle these two effects in the VO2 voltage driven IMT process [98,99].

### 3. Result and discussion

To minimize the EM coupling between the adjacent meta-particles, the VBCM structure comprises of $N \times N$ array of $M \times M$ identical subwavelength unit-cells that constructs the so-called super-unit cell or lattices. The length of one lattice is $MP$ and the length of whole VBCM is equals to $NMP$. The metasurface is assumed to be illuminated by a y polarized normal incident plane wave throughout the paper. From the well-known antenna theory, the far-field scattering pattern function of the VBCM can be expressed by [12]:

$$E_{scat}(\theta.\varphi) = E_{elem}(\theta.\varphi) \times F(\theta.\varphi) \qquad (2)$$

$$F(\theta.\varphi) = \sum_{m=1}^{N}\sum_{n=1}^{N} a_{mn} exp[-j\{\phi(m.n) + KPsin\theta[(m - {}^{1}/_{2})cos\varphi + (n - {}^{1}/_{2})sin\varphi]\}] \qquad (3)$$

in the above equations, θ and φ are the elevation and azimuth angles of the desired direction, $P$ demonstrates the period of lattices along both x and y directions, $a_{mn}$ and $\phi_{mn}$ are the demonstrator of reflection amplitude and phase of each lattices respectively and $K = 2\pi/\lambda$ where $\lambda$ is a working wavelength. Establishing a 2D inverse fast Fourier transform (2D-IFFT) to accelerate the calculation make this formula beneficial for the prediction of scattering patterns caused by different coding sequences. The VBCM structure occupied with $8 \times 8$ array of $6 \times 6$ identical unit-cells throughout this study. We begin with the simplest example driven by uniform coding sequence of (0000…/0000…) to generate a reflection beam toward the normal direction. The 3D far-field pattern of such a perfect electric conductor (PEC) behavior is depicted in Fig. 3(a). In the two next layout, only digital elements of "00" and "10" with 180° phase difference are adopted. By distributing these two digital elements in alternate rows or columns, the reflected wave is split into two symmetrically oriented scattered beams ($\theta = 34°.\varphi = 0°.180°$) with a null in the boresight direction governed by generalized Snell's law[100] as shown in Fig. 3(b). By arranging the above two digital elements in a chessboard configuration (see Fig. 3(c)), the incoming energy redirected into four scattered main beams of ($\theta = 53°. \varphi = 45°. 135°. 225°. 315°$) at 0.44THz that have an excellent conformity with the theoretical predictions. According to generalized Snell's law, with increasing the observation frequency, the reflection beams come closer to the boresight direction [78].

### 3.1. Phase-gradient Coding Sequence: anomalous reflection

1-bit coding metasurface is not capable of producing an arbitrary tailored single reflected beam as an exciting example of THz wavefront engineering. By extending the concept, our 2-bit VBCM structure has greater freedom to manipulate EM waves for instance creating adjustable pre-defined anomalous reflection driven by phase-gradient coding sequence. Suppose normally incident wave are illuminating the VBCM structure with phase gradient coding sequences of [00,01,10,11…/00,01,10,11…] as shown in Fig. 4(a). According to the generalized Snell's law, the reflection angles $\theta_r$ and $\varphi_r$ can be written as:

$$\theta_r = arcsin\frac{\lambda}{4MP} \qquad (4)$$

$$\varphi_r = arctan\frac{\Delta\phi_y}{\Delta\phi_x}\frac{D_x}{D_y} \qquad (5)$$

Where $\Delta\phi_y$ and $\Delta\phi_x$ are the phase differences of super-unit-cells along the x and y directions, respectively and $D_x = D_y = P$. In the current phase gradient prototype $\Delta\phi_x = \pi/2$ and $\Delta\phi_y = 0$ which makes the incident THz wavefront reflected into arbitrary pre-determined oblique angles of ($\theta_r = 16°. \varphi_r = 180°$) which are in good agreement with theoretical predictions as depicted in Fig. 4(a). To further demonstrate the flexibility of our proposed structure and diverting an impinging wave into a new pre-determined oblique angles of ($\theta_r = 24°. \varphi_r = 45°$), a 2-bit VBCM is elaborately encoded exploiting a phase gradient coding sequences of [00,01,10,11…/01,10,11,00…/10,11,00,01…/11,00,01,10…] as shown in Fig. 4(b) leading to $\Delta\phi_x = \Delta\phi_y = \pi/2$. The 3D scattered far-field pattern of such encoded metasurface is depicted in Fig 4(b).

Obviously in anomalous reflection phenomena, by purposefully arranging the coding sequences, the reflected beam can be directed toward a pre-determined azimuth angle in each of four quadrants of ($\varphi_r = 45°. 135°. 225°. 315°$) and furthermore at the same frequency, the generated beam elevation angle can be shifted by changing the dimension of lattices of the VBCM structure and beam steering functionality could be envisioned. To validate the concept, the phase gradient VBCMs employed with different size lattices corresponding to $M = 3. 6. 7$ are analyzed by the full wave simulation (see Fig. 5) and a good agreement between numerical simulations and theoretical predictions confirm the validity of the presented beam steering analysis in a real-time ultrafast VBCM structure. These examples clearly illustrate that the instantaneous access to an anomalous reflection with outgoing directions is simply realized with our designed VBCM structure. Eventually, by varying the biasing voltage controlled by an FPGA platform, a single beam that can dynamically altered in any desired angles with ultrafast switching time is provided with our presented structure which has fascinating functionality to implement in various applications such as ultrafast THz wireless communications and tracking systems.

### 3.2. multi-beam generation

Recently, it has been revealed that when two different coding patterns are added together by means of the superposition theorem [101], combined coding pattern will perform both functionalities simultaneously aid to reach a metasurface with several missions such as multi-beam generation. Two different layouts will be anticipated here to picture the multi-mission capability of the VBCM device. Let us consider two metasurfaces driven by gradient coding sequences along different directions to generate two single beams toward ($\theta_r = 25°. \varphi_r = 45°$) and ($\theta_r = 25°. \varphi_r = 315°$) respectively. We adopted the superposition theorem here to design a 2-bit VBCM that redirect the incident THz wavefront from the normal direction into two reflected beams with ($\theta_r = 25°. \varphi_r = 45°. 315°$) driven by a coding sequences generated by:

$$\angle[e^{j\varphi_1} + e^{j\varphi_2}] = e^{j\varphi_0} \tag{6}$$

in which $\varphi_1$ and $\varphi_2$ are arguments of the two primary complex codes and $\varphi_0$ represents the argument of superimposed complex code. The 2D phase map of two independent phase gradient metasurface and the final required phase profile of the mixed metasurface to generate multi-beams is depicted in Fig. 6(a) and the 3D scattering pattern of superimposed VBCM is shown in Fig. 6(b).

The next example is dedicated to adopting superposition theorem in order to add two different gradient coding patterns with distinct elevation angles. Before we delve into the full wave simulations, it should be noted that when multiple independent pencil beams with different elevation angles are added together, the superimposed coding metasurface generates multiple beams with asymmetric power ratio levels. More recently, by revisiting the addition theorem in the metasurface, our team provided a generalized version of the superposition theorem to estimate the exact amount of power ratio of the multiple beams [102]. Considering the cosine function as the element factor of the metasurface particles, the amount of power distribution of two asymmetric pencil beams can be estimated as follows:

$$\frac{P(\theta_2)}{P(\theta_1)} \propto \left[\frac{\cos(\theta_2)}{\cos(\theta_1)}\right]^2 \tag{7}$$

which means that in our designed VBCM structure with two asymmetrically oriented scattering beams, the arbitrary oriented pencil beam with higher elevation angle ($\theta$) carring lower power intensity. Let us consider, two metasurfaces with phase gradient coding sequences one of which makes the incident wave reflect at oblique angles of ($\theta_r = 17°. \varphi_r = 180°$) while the other generates single anomalously scattered beam toward ($\theta_r = 25°. \varphi_r = 45°$). Exploiting the same design approach results a VBCM structure that has incorporated both aforesaid gradient codes satisfactorily generates two asymmetrically oriented reflected beams along the pre-determined directions as shown in Fig. 6(d). Inspired by the addition theorem, these examples proved the correct functioning of our proposed VBCM structure to create symmetric/asymmetric multiple beams in any desired directions which is dynamically interchangeable.

### 3.3. Spiral Coding Sequence: vortex beam generation

**3.3(a). Single vortex wavefront carrying different OAM modes**

Since being discovered in 1992 [103], vortex beams carrying orbital angular momentum (OAM) has experienced increasing levels of attention for its potential opportunities in high speed communication [104], fast imaging [105], optical manipulation [106], etc. Advantages such as improving the channel capacity without increasing the bandwidth and orthogonality of different topological charges [107] have prompted researchers to present various novel method for generating vortex beams from spiral phase plates [108] to antenna arrays [109]. However, there are still critical bottlenecks which are waiting for the solutions so that only a few reports involve

THz vortex beam realization. Furthermore, future progress in classical and quantum systems require rapidly switch between different OAM modes and at the present time they are suffering from lack of ultrafast reconfigurable THz device in order to switch between different OAM modes in a real time manner.

In order to address the aforesaid restrictions and to further clarify the versatility of our proposed ultrafast tunable VBCM structure, four THz vortex wavefront with different OAM modes ( $l = 2. -2. +1. -1$) are envisaged. In each case, the coding metasurface is encoded by a spiral coding sequences generated by [79]:

$$\varphi(x.y) = l \times \arctan\left(\frac{y}{x}\right) \tag{8}$$

In another words, after dividing coding metasurface into N equal segments with phase differences $\Delta\varphi$ of the neighboring segments, relation between OAM topological charges and the number of segments for our 2-bit coding metasurface yields to:

$$N \cdot \Delta\varphi = 2\pi l \xRightarrow{\Delta\varphi=\frac{\pi}{2}} N = 4l. \tag{9}$$

Toward this aim, in order to generate OAM beams with $l = \pm 1$ and $l = \pm 2$, we devide the VBCM structure into four and eight segments with phase shift ranging from 0 to $2\pi$ and 0 to $4\pi$ respectively. The instant electric field intensities of these four configuration when the observation plane is set as 170 $\mu m$ away from the center of VBCM structure with an area of 4800 $\mu m \times$ 4800 $\mu m$, shown in Fig. 7(a)-(d). 3D far-field scattering patterns (see Fig. 7(e)-(f)), demonstrate that the VBCM can successfully generate vortex beams carrying OAM ($l = 2. -2$) at 0.44 THz. According to Fig 7, a typical doughnut-like intensity profile with an amplitude null in the center (15 dB lower than the annular high-intensity region) satisfies the far-field feature of OAM beams. The capability of ultrafast switching between different topological charges in a real time manner is provided by our VBCM structure that can empower dramatic advances in wideband OAM based multi-user system where the beams topological charges identifies the routing.

### 3.3(b). Arbitrarily-oriented multiple vortex beams

As a fascinating applications in THz OAM-based MIMO systems, generating oriented vortex beam at a predetermined direction which is dynamically interchangeable can solve the crucial challenges of this research area. Convolution operation as a simple yet wonderful solution has armed a platform to produce obliquely directed OAM vortex wavefront. Regarding the Fourier transform relation between the coding patterns and its far-field scattering patterns, the scattering pattern shift functionality can be achieved by the means of convolution theorem [110] by adding spiral phase distribution with gradient coding sequences. We start with a simple gradient coding sequence of [00,01,10,11…/00,01,10,11…] which generates single anomalously scattered beam toward ($\theta_r = 17°. \varphi_r = 180°$). By multiplying spiral phase coding pattern ($l = 1$) by the above gradient coding sequence, a tilted vortex scattering pattern is generated in our desired direction (see Fig. 8(a)-(d)). As the next layout, a phase gradient coding sequences of $M_1 = $ [00,01,10,11…/01,10,11,00…/10,11,00,01…/11,00,01,10…] has added to elaborately encoded four segments metasurface with rotated phase distribution ($l = 1$) of $M_2$ to generate an obliquely directed OAM vortex wavefront in predetermined direction ( $M_3 = M_1 + M_2$ ). The 2D reflection phases map and 3D far-field scattering pattern of such encoded metasurface ($M_3$) is depicted in Fig. 8(e)-(i). As can be deduced from Fig. 8(i), the rotation angle of generated arbitrarily-oriented OAM beam with $l = 1$, which is dictated by the phase gradient coding sequence has a good conformity with our theoretical predictions based on equations 4 and 5.

To further specify the ability of presented structure to generate multiple vortex beams, we add the phase distribution of suitably programed VBCM to create vortex beam carrying OAM mode $l = 1$ with stripped and chessboard configurations respectively. Fig. 9 display the simulated scattering patterns of such encoded metasurfaces after adopting convolution operation. As can be deduced from Figure 9(c) and (f), the mixed coding patterns satisfactorily generates two and four symmetrically oriented reflected vortex beams pointing at pre-determined directions of ($\theta_r = 24°. \phi_r = 0°. 180°$) and ($\theta_r = 28°. \phi_r = 45°. 135°. 225°. 315°$) respectively that are very close to our theoretical predictions based on equations 4 and 5. In these two examples we choose $M = 8$ and $M = 10$ and VBCM structure is occupied with $64 \times 64$ and $60 \times 60$ unitcells respectively. Inspired by the convolution theorem, these examples revealed the competence of our VBCM structure to create arbitrarily-oriented multiple vortex beams in pre-demined directions with different topological charges in a real time manner.

### 3.3(c). Focused vortex beam

The versatility of our designed VBCM also has equipped a platform to focus the incident THz wavefront into a pre-determined point. The focal length ($z_{focal}$) can be dynamically altered by suitably changed the biasing system of VBCM driven by a parabola phase distribution along the radial direction. Furthermore, our proposed structure provides the ability to obtain several focused vortex beams with ultrafast switching time between different topological charges and focal lengths. To engineer this feature of the work, the phase profile of such encoded VBCM must involves both spiral and parabolic phase distributions simultaneously which can be expressed by [111]:

$$\varphi(x.y) = l \times \arctan\left(\frac{y}{x}\right) + \frac{2\pi}{\lambda}\left(\sqrt{x^2 + y^2 + z_{focal}^2} - z_{focal}\right) \quad (10)$$

To validate the concept, four focused wavefront carrying OAM with diverse topological charges and focal lengths of $(l.z_{focal}) = (-1.800\mu m)$, $(l.z_{focal}) = (+1.1100\mu m)$, $(l.z_{focal}) = (+2.900\mu m)$ and $(l.z_{focal}) = (-2.1200\mu m)$ have been designed and exemplified in Fig. 10. In each case, the required 2D spiral-parabola phase map for realizing the corresponding digital states have been pictured in Fig. 10(e)-(h) respectively. Eventually, illuminating by a y polarized normal incident plane wave, the focused-vortex-generating VBCM structures are build-up and the simulated normalized electric near-field intensities in the sampling planes at the corresponding focal lengths (see Fig. 10(a)-(d)) demonstrate the capability and flexibility of the designed VBCM to generate THz focused vortex beams with different OAM modes and focal lengths which is dynamically switchable.

### 4. Potential fabrication procedure of the designed structure

In this section we furnish a brief presentation on the current fabrication technologies [21,112] for fabricating our proposed VO2-based meta atom structure. This process can follow the steps below (see Fig. 11): (a) photoresist is spin coated and deposited on a substrate using photolithography to form the pattered VO2 bricks; (b) a desired sample pattern is transferred to photoresist by writing the pattern with an electron beam lithography; (c) $1\mu m$ thick VO2 layer is then prepared on 38 $\mu m$ c-type sapphire substrate using magneton sputtered technique; (d) lift-off process in order to dissolve the photoresist and leave behind the film only in the patterned area is proposed and the remaining patterned VO2 bricks is annealed at $\sim 450^o C$; (e) finally gold layer is deposited on the backside of the substrate as a ground layer. Following above steps, one can envision a practical fabrication within the scope of the current fabrication technologies. Eventually, two thin Au pads is deposited on each VO2 bricks and serve as the local bias elements. By applying proper biasing voltage through the wires directly connected to VO2 films [41], reconfigurability of the VBCM structure will be feasible. The detailed information about the possibility of triggering the MIT phase transition is reported in [41]. According to [113,114], very thin biasing lines with respect to the operating wavelength (analogous to this paper), will not perturb the far field pattern and therefore have negligible impact on the macroscopic behavior of our designed VBCM and the individual response of meta atoms.

## 5.Conclusion

The necessity of integrating multiple diversified functionalities in a single structure at THz frequencies sparks our curiosity to design a re-programmable multi-mission coding metasurface incorporated with phase-changes materials. In this paper, for the first time we have proposed a novel reconfigurable metasurface to dynamically manipulate THz wavefront by utilizing the insulator to metal transition in VO2. VO2 thin film exhibits ultrafast switching time which is of great practical significance in ultrafast THz communication. The versatility of our proposed VBCM structure was successfully clarified with different illustrative examples from anomalous reflection to focused vortex beam generation. Furthermore, symmetric/asymmetric multiple beams along the pre-determined directions have also been realized by applying superposition operation. By encoding the VBCM by a spiral-like and spiral-parabola-like coding sequences, single vortex beam carrying OAM and focused vortex beam were satisfactorily generated respectively. Additionally, by adopting convolution operation, arbitrarily-oriented multiple vortex beams in pre-demined directions with different topological charges in a real time manner was realized with our elaborately designed VBCM. Our ultrafast reconfigurable THz device meets well the THz future industrial demands which require rapidly switch between different OAM modes. In the last section we demonstrated the feasibility of fabricating the VBCM within the realm of the current fabrication technologies and this simple yet fruitful structure holds great potential for dynamically THz wavefront engineering and can enable advanced applications such as high data rate wireless communication and ultra-massive MIMO communication.

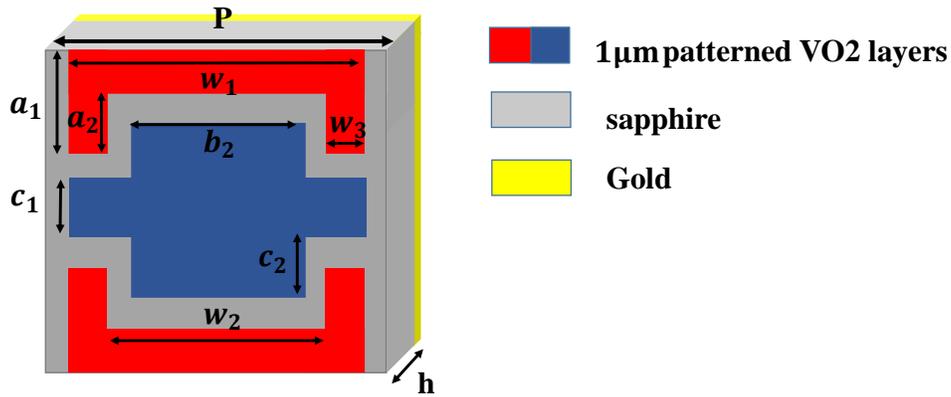

Figure 1. The front view of the employed VO2 based meta-atom of the proposed switchable multifunctional THz coding metasurface.

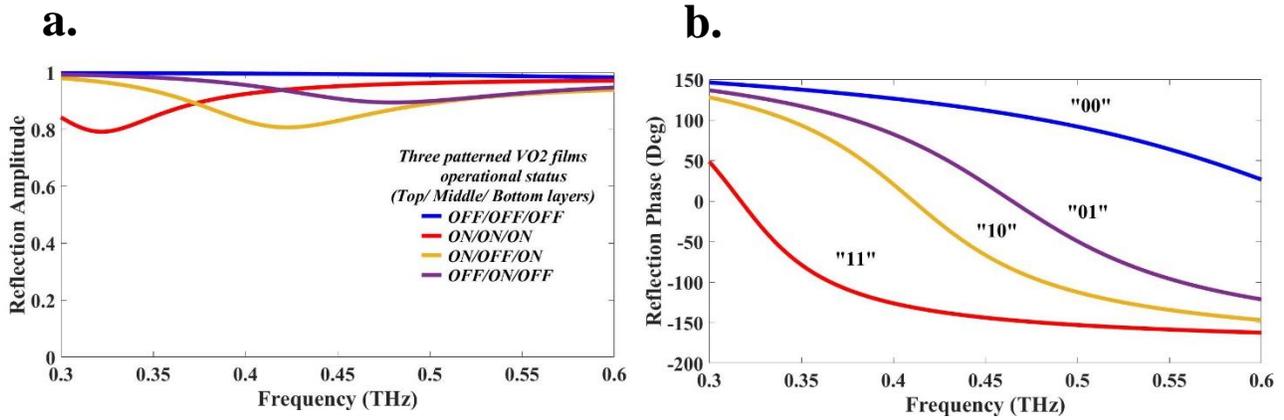

Figure 2. Simulated reflection spectra (a) amplitude and (b) phase of VO2 based meta-atom under illuminating y-polarized normal incident plane wave. It can be found that the perfect 2-bit characteristics (phase step of 90°) occurs at 0.44 THz as a working frequency.

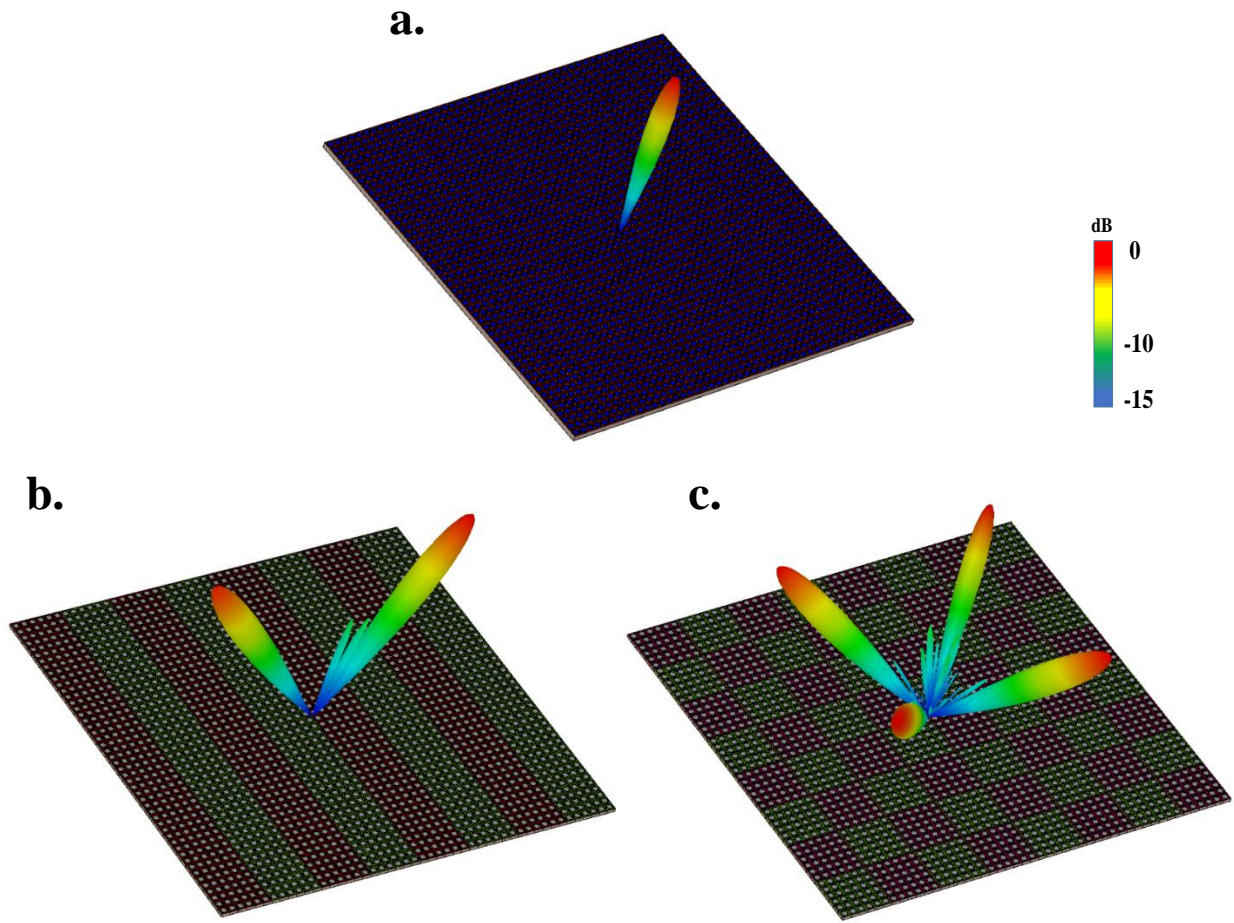

Figure 3. The simulated 3D far-field pattern results at 0.44 THz by adopting only digital elements of "00" and "10" with 180° phase difference (a) perfect electric conductor behavior (b) VBCM structure encoded by [00,10,00,10…/ 00,10,00,10…] which splits the incident THz wavefront into two symmetrically oriented scattered beams (c) VBCM driven by [00,10,00,10…/10,00,10,00…] which redirected the incoming energy into four scattered main beams, which are in a excellent conformity with the analytical predictions.

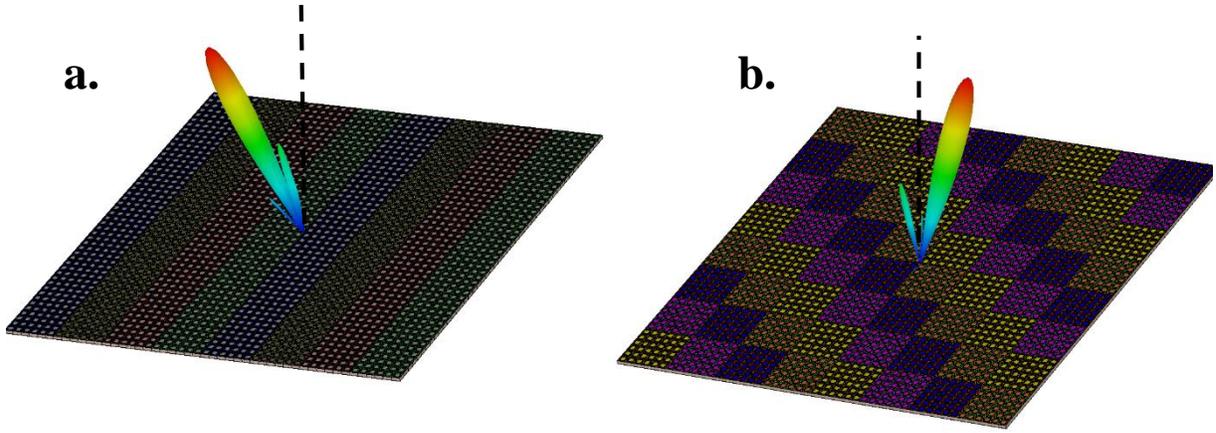

Figure 4: 3D far-field scattering pattern of anomalous reflection toward (a) ($\theta_r = 16°. \phi_r = 180°$) (b) ($\theta_r = 24°. \phi_r = 45°$) when 2-bit VBCM is encoded by phase gradient coding sequences of [00,01,10,11…/00,01,10,11…] and [00,01,10,11…/01,10,11,00…/10,11,00,01…/11,00,01,10…] respectively.

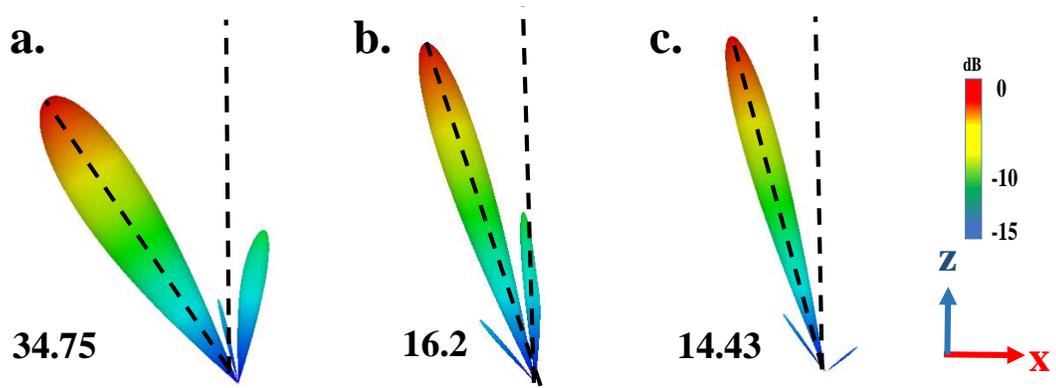

Figure 5: Schematic of 2-bit VBCM driven by [00,01,10,11…/00,01,10,11…] and its 3D far-field scattering patterns for (a) M=3, (b) M=6, (c) M=7. Good agreement between full wave simulations and theoretical predictions confirm the dynamic beam steering feature of our proposed 2-bit VBCM structure.

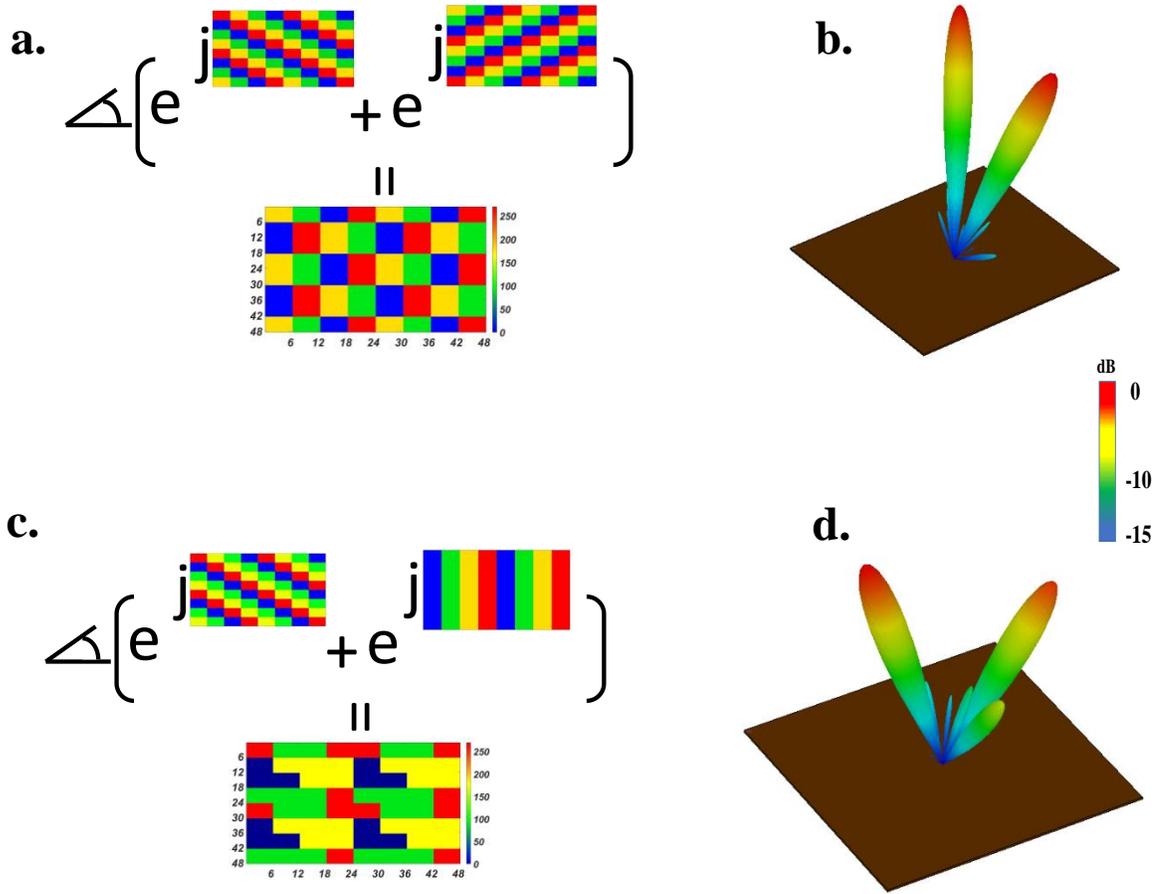

Figure 6: Intuitive display for superposition theorem when two metasurface with gradient coding sequences are added together and two single beams toward (b) ($\theta_r = 25°. \phi_r = 45°. 315°$) and (d) ($\theta_r = 17°. \phi_r = 180°$). ($\theta_r = 25°. \phi_r = 45°$) are generated successfully. (a) and (c) the Initial phase distribution of the metasurfaces before additional operation and the final phase distribution of the mixed metasurface.

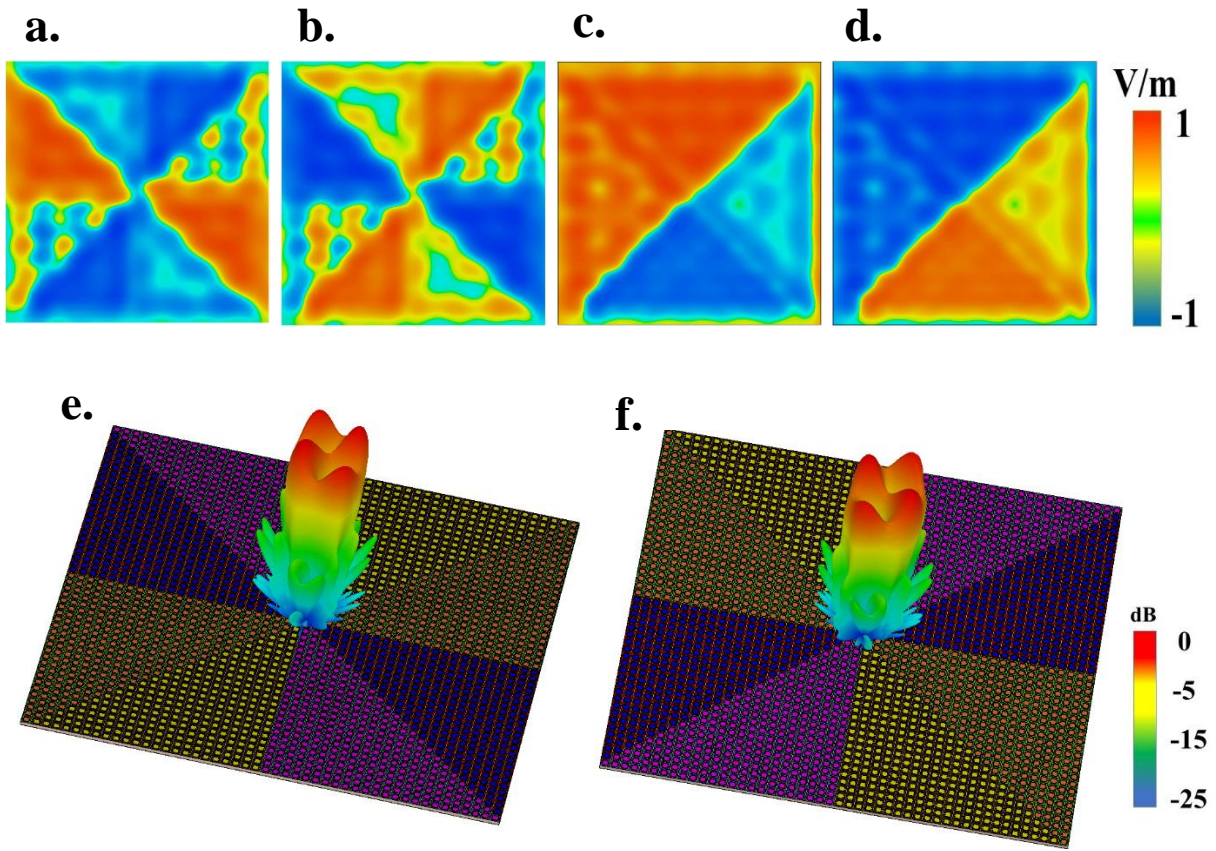

Figure 7: Simulated y-component of electric field distribution of generated vortex beams with the topological charges of (a) $l = +2$; (b) $l = -2$; (c) $l = +1$ and (d) $l = -1$. 3D far-field scattering patterns of VBCM structure after dividing the coding mrtasurface into eight segments driven by spiral coding sequences with topological charges of (e) $l = +2$ and (f) $l = -2$.

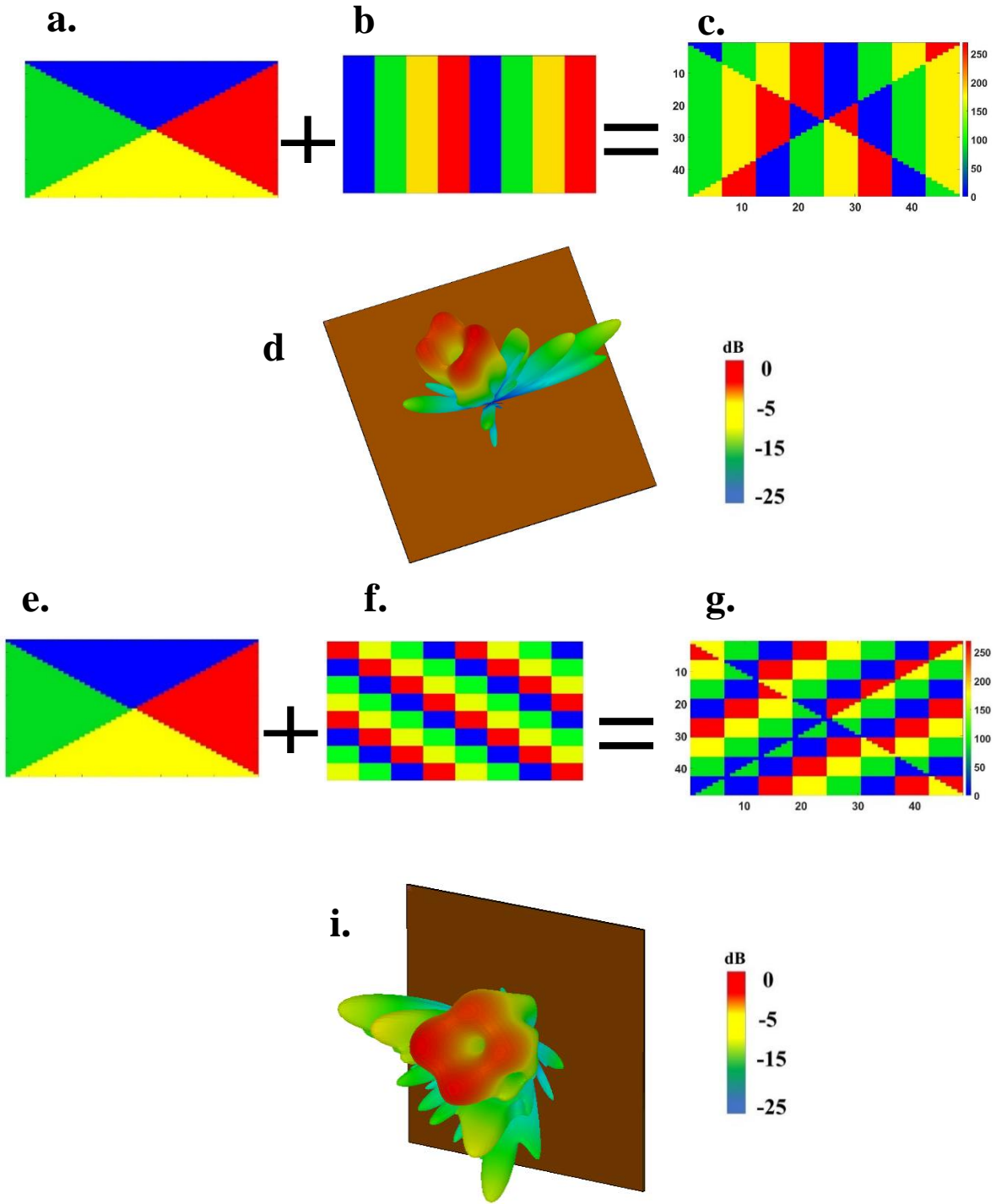

Figure 8: Intuitive display for convolution operation when (a) spiral coding sequence is added with (b) gradient coding sequence of [00,01,10,11…/00,01,10,11…] to generate (c) oriented-vortex beams toward ($\theta_r = 17°. \phi_r = 180°$) and (d) the 3D far-field scattering pattern of such encoded VBCM structure. By multiplying (e) spiral coding sequence with (f) gradient coding sequence of [00,01,10,11…/01,10,11,00…/10,11,00,01…/11,00,01,10…], (g)-(i) a tilted-vortex beam is generated toward predetermined direction of ($\theta_r = 25°. \phi_r = 45°$) .

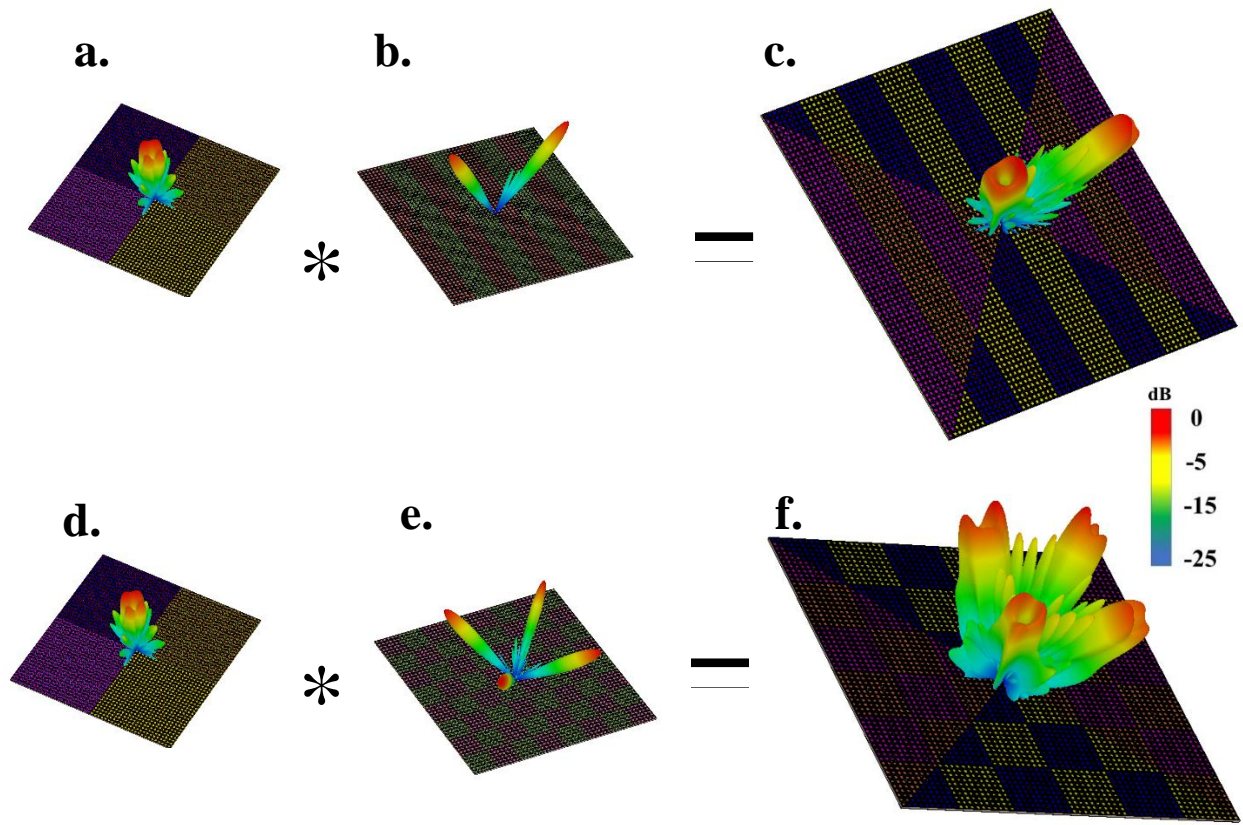

Figure 9: Intuitive presentation of convolution operation on the metasurface levels. (c)-(f) 3D far-field scattering patterns of VBCM structure after multiplying vortex beam with stripped and chessboard configurations respectively. In these two examples we choose (c) $M = 8$ and (f) $M = 10$ and VBCM structure is occupied with $64 \times 64$ and $60 \times 60$ unitcells respectively.

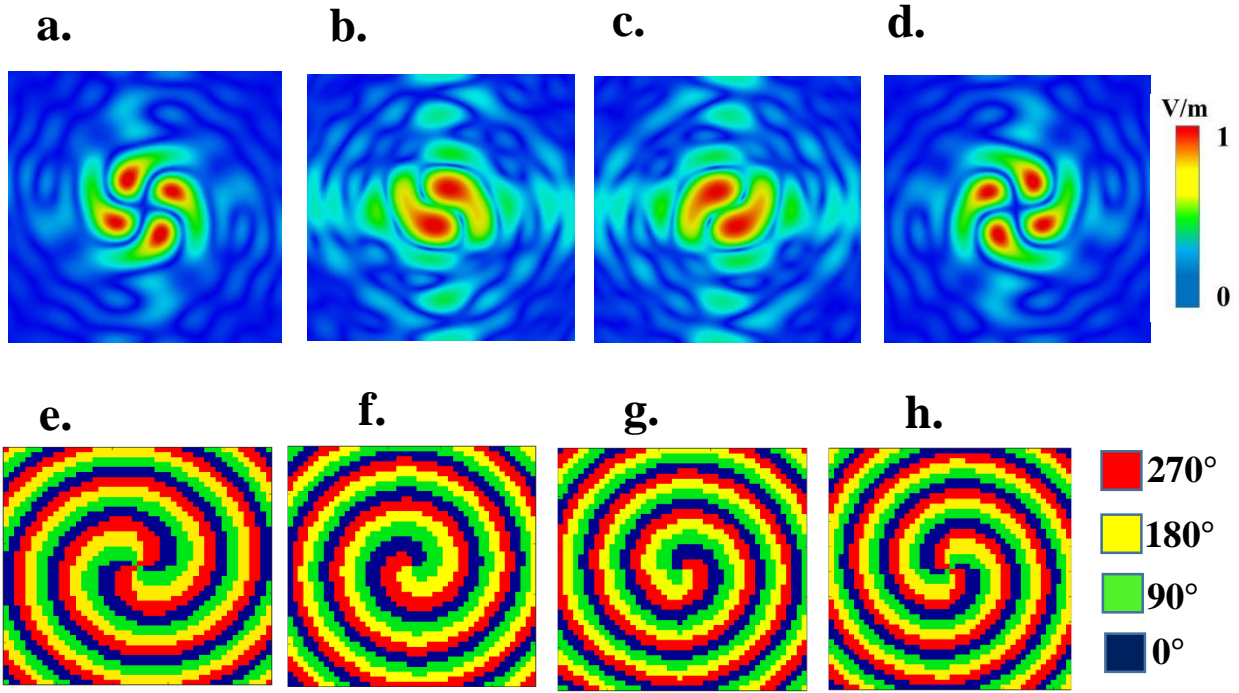

Figure 10: Simulated normalized electric near-field intensities in the sampling planes in xoy plane at different focal distances of (a) $1200\mu m$ (b) $800\mu m$ (c) $1100\mu m$ (d) $900\mu m$ corresponding to $l = -2, -1, +1, +2$ respectively. (e) –(f) The required 2D spiral-parabola phase map of VBCM structure to generate different focused vortex beams.

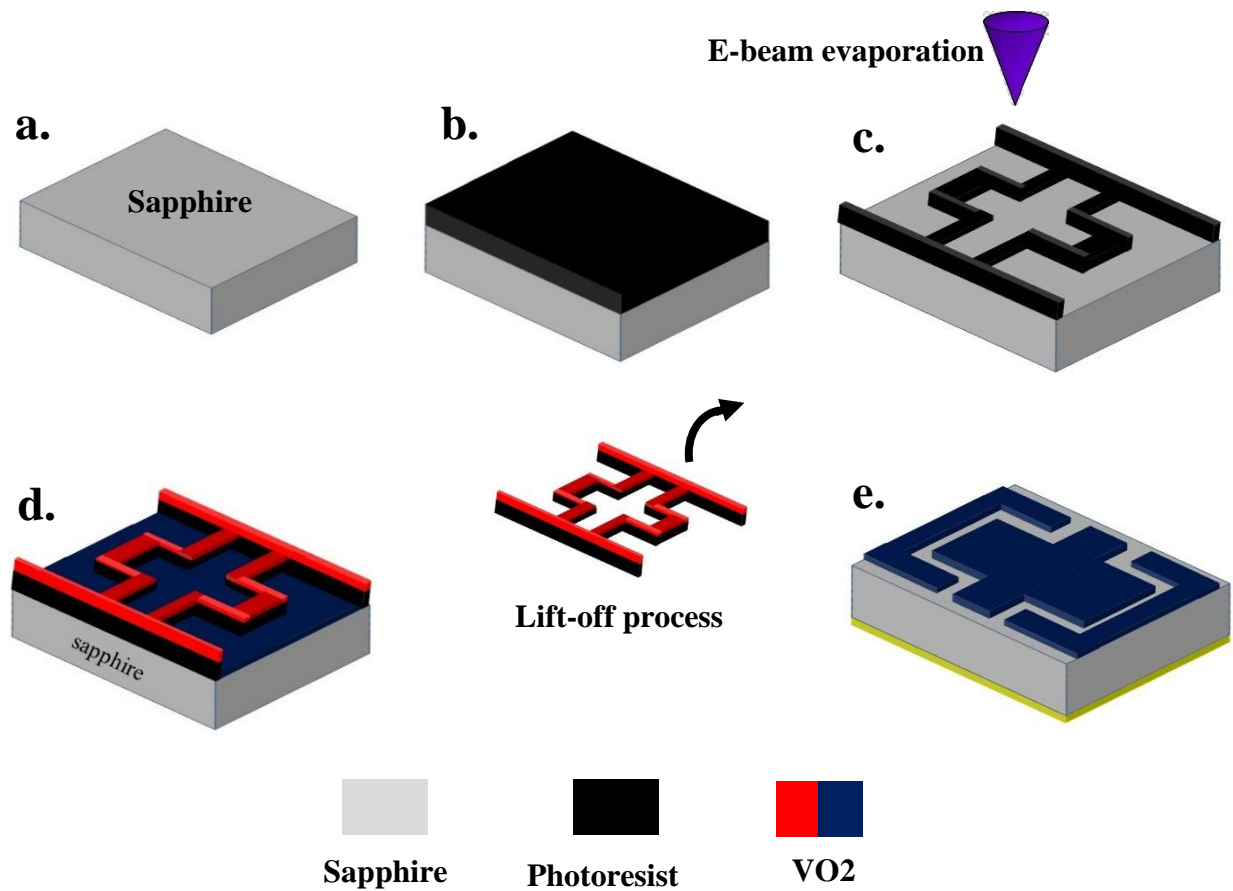

Figure 11: Potential fabrication procedure of the designed structure: (a) 38μm thick c-cut sapphire is prepared; (b) Photoresist is spin coated and deposited on a substrate using photolithography; (c) Photoresist is patterned with an electron beam lithography; (d) 1μm thick VO2 layer is deposited using magneton sputtered technique; (e) Final VBCM structure after lift-off process to dissolve the photoresist and the remaining patterned VO2 bricks is annealed at ~450°$C$. Note: VO2 is shown in two different colors (red and blue) just for better visualization.